\documentclass[aps, prb, twocolumn, superscriptaddress, showpacs]{revtex4}

\usepackage{amssymb}
\usepackage{amsmath}
\usepackage{bm}
\usepackage{graphicx}
\usepackage{dcolumn}
\usepackage{longtable}

\newcommand{\kNCS}{$\kappa$-(BEDT-TTF)$_{2}$Cu(NCS)$_{2}$ }
\newcommand{\kBr}{$\kappa$-(BEDT-TTF)$_{2}$Cu[N(CN)$_{2}$]Br }

\begin{document}

\title{Microscopic observation of superconducting fluctuations in \kBr by $^{13}$C NMR spectroscopy}
\author{T.~Kobayashi}
\author{Y.~Ihara}
\altaffiliation{yihara@sci.phys.hokudai.ac.jp}
\author{Y.~Saito}
\author{A.~Kawamoto}
\affiliation{Department of Condensed Matter Physics, Graduate School of Science, Hokkaido University, Sapporo 060-0810, Japan}
\date{\today}

\begin{abstract}
We performed $^{13}$C-NMR experiment and measured spin-lattice relaxation rate divided by temperature $1/T_{1}T$ near the 
superconducting (SC) transition temperature $T_{c}$ 
in \kBr ($\kappa$-Br salt), and \kNCS ($\kappa$-NCS salt). 
We observed the reduction of $1/T_{1}T$ starting at the temperature higher than $T_c$ in $\kappa$-Br salt. 
Microscopic observation of quasi-particle density of states in the fluctuating SC state 
revealed the effects of short-range Cooper pairs induced in the normal state to the quasi-particle density of states. 
We also performed systematic measurements 
in the fields both parallel and perpendicular to the conduction plane in $\kappa$-Br and $\kappa$-NCS salts, 
and confirmed that the reduction of $1/T_{1}T$ above $T_{c}$ is observed only in
$\kappa$-Br salt regardless of the external field orientation.

\end{abstract}

\pacs{74.25.nj, 74.70.Kn, 75.25.Dk}
\maketitle

\section{INTRODUCTION}

Near the second order phase transition, thermal and quantum fluctuations induce a short-range coherence even above the transition temperature. 
The universality of physical quantities in this critical regime has been intensively studied for various types of magnetic transitions \cite{millis-PRB48}. 
For the second-order superconducting (SC) transition, 
the short-range coherence of SC order parameter should also be induced in the normal state, 
where the thermal average of order parameter does not possess the finite value. 
The anomaly accompanied with fluctuating SC order parameter should be investigated to 
proceed universal understanding of the second-order phase transition. 
The prominent effect of fluctuating SC order parameter was observed by the Nernst effect measurement 
on cuprate superconductor \cite{xu-nature}, 
in which a vortex-like signal was detected in the critical regime above the SC transition temperature $T_{c}$. 
To reveal the mechanism for the fluctuating SC order parameter to affect the bulk properties, 
both theoretical and experimental studies have been carried out on the high-$T_{c}$ cuprate superconductors \cite{larkin, ri-PRB50, carretta-PRB61, mitrovic-PRL82}. 
However, enhanced magnetic fluctuations, which originate from the antiferromagnetic transition near the SC phase, 
contaminate the pure effect of fluctuating SC order parameter. 
In order to investigate the critical fluctuation of SC order parameter, 
we should examine a superconductor that shows SC transition in the conventional Fermi liquid state. 

The observation of vortex-like signal by Nernst effect measurement in an organic superconductor \kBr 
(BEDT-TTF: bis(ethylenedithio)tetrathiafulvalene) 
leads us to study the fluctuating SC state in a series of $\kappa$-(BEDT-TTF)$_{2}X$ salts \cite{nam-nature449}, 
because in the organic superconductors the conventional Fermi liquid state is established when superconductivity sets in. 
Besides, the effects associated with superconductivity can be controlled by magnetic fields, 
as the upper critical field $B_{c2}^{\perp}\simeq 10$ T is accessible with conventional SC magnets \cite{mayaffre-PRL75, kwok-PRB42}. 
The precursor of SC transition above $T_{c}$ was detected also by magnetic torque, 
and magnetization measurements \cite{tsuchiya-PRB85, lang-PRB49, uehara-JPSJ82}. 
These experiments probed the vortices generated by the short-range Cooper pairs in the normal state, 
and confirmed that fluctuating SC state is realized in \kBr salt. 
Since the fluctuating SC state has been established, 
we should explore the variation of quasi-particle density of states (DOS) caused by the fluctuating SC order parameter, 
which is essential for the microscopic understanding of critical SC fluctuation \cite{randeria-PRB50}, 
but cannot be observed by the bulk measurements because of the large magnetization from vortices.  

The NMR spectroscopy enables us to measure directly the quasi-particle DOS when the system is in the Fermi liquid state. 
The previous nuclear spin-lattice relaxation rate $1/T_{1}$ measurement on cuprate superconductors 
could not observe separately the effect of SC fluctuations \cite{carretta-PRB61,mitrovic-PRL82}, 
because the strong spin fluctuations near antiferromagnetic phase violate the conventional Fermi liquid state. 
In the organic superconductors, as the spin fluctuations are weak enough to stabilize the Fermi liquid state above $T_{c}$, 
NMR experiment should be performed on $\kappa$-(BEDT-TTF)$_{2}X$ salts.

The $\kappa$-(BEDT-TTF)$_2X$ family salts are quasi-two dimensional conductors. 
When $X=$Cu(NCS)$_{2}$ ($\kappa$-NCS salt), superconductivity is observed below $T_{c}=9.4$~K \cite{urayama-chemlett17}, and 
when $X=$Cu[N(CN)$_{2}$]Cl, antiferromagnetic transition occurs at $T_N=22$~K \cite{welp-physicaB186-188}. 
Another superconducting salt with $X=$Cu[N(CN)$_{2}$]Br ($\kappa$-Br salt) locates very close to the antiferromagnetic phase, 
but the typical Fermi liquid behavior, namely temperature independent bulk susceptibility and $T^{2}$ dependence of resistivity, 
was clearly observed below $20$ K \cite{kawamoto-PRL74, strack-PRB72}. 
The deuteration to $\kappa$-Br salt brings the electronic state toward the antiferromagnetic phase, and 
enhanced spin fluctuations in the deuterated $\kappa$-Br salt violates the Fermi liquid state \cite{kawamoto-JACS120}. 
The deuterated salts are referred to as $\kappa$-$d[n,n]$-Br ($n = 0 \sim4$) depending of the numbers of 
substituted hydrogen at each ethylene group of BEDT-TTF molecule. 
The $^{13}$C NMR experiment on $\kappa$-$d[4,4]$-Br salt observed the reduction of $1/T_{1}T$ much above $T_{c}$ \cite{miyagawa-PRL89}. 
This anomalous reduction of $1/T_{1}T$ can be interpreted by either magnetic or SC fluctuations, just as in the case for cuprate superconductors. 
Therefore, we measured $1/T_{1}T$ on non-deuterated $\kappa$-$d[0,0]$-Br and $\kappa$-NCS salts, 
which demonstrate the Fermi liquid state at $T_{c}$, 
and investigated the pure effect of fluctuation SC order parameter. 
The Nernst effect measurements detected the vortex-like signal above $T_{c}$ in $\kappa$-Br salt, 
and no such signal in $\kappa$-NCS salt \cite{nam-nature449}. 
The $X$ anion dependence of SC criticality was interpreted as the difference in the proximity to antiferromagnetic phase. 
Nam {\it et al.} proposed that fluctuation in the phase of SC order parameter would be induced 
when the quasi particles are nearly localized in the vicinity of antiferromagnetic insulator phase. 
On the other hand, magnetic torque measurements concluded that comparable SC fluctuation effect was 
observed in $\kappa$-Br and $\kappa$-NCS salts \cite{tsuchiya-PRB85, tsuchiya-JPSJ82}. 
This inconsistency can be due to the difference in the external field direction. 
The external field was applied perpendicular to the conduction plane for the Nernst effect measurement, 
whereas parallel to the conduction plane for the magnetic torque measurement. 
As the experimental constraint determines the field orientation for these techniques, 
other experiments that can be conducted in both parallel and perpendicular fields are required. 
In the present $^{13}$C NMR experiment on $\kappa$-Br and $\kappa$-NCS salts, 
we measured $1/T_{1}T$ in the magnetic field both parallel and perpendicular to the conduction plane.

\section{Experimental}

The single-crystal $\kappa$-Br and $\kappa$-NCS salts were grown using electrochemical reaction. 
The typical dimension is $1.9\times 1.5\times 0.6 $ mm$^3$. 
We utilized the $^{13}$C enriched BEDT-TTF molecule, 
for which single side of the central C=C bond was enriched with $^{13}$C \cite{yamashita-synthmet133-134}. 
This asymmetric molecule eliminates the NMR peak splitting due to the nuclear spin-spin coupling at the C=C bond (Pake doublet) \cite{pake-JCP16}. 
We determined the mean-field $T_{c}$ of our samples from magnetization measurement at low field as 
$T_{c}=11.4$ K for $\kappa$-Br, $T_{c}=9.3$ K for $\kappa$-NCS, 
which are consistent with the previous studies, and evidence the high quality of our samples. 
\begin{figure}[tbp]
\begin{center}
\includegraphics[width=7cm]{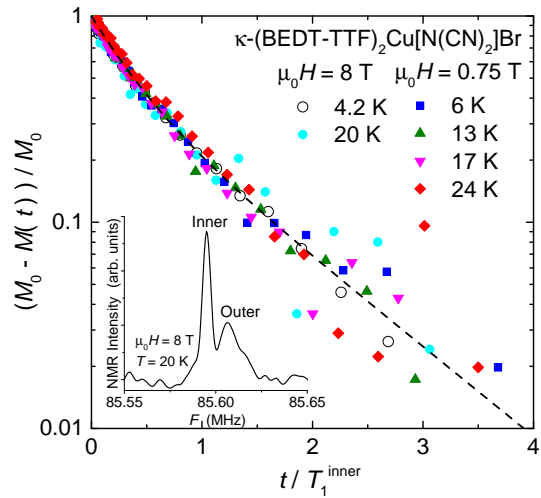}
\end{center}
\caption{ 
Recovery profile of nuclear magnetization obtained in wide range of temperatures and magnetic fields. 
The horizontal axis was scaled by $T_{1}^{\rm inner}$ determined at each temperature. 
The universal behavior was observed both in the SC and normal states, 
which confirms that the obtained $T_{1}$ reflects intrinsic character of quasi-particle density of states. 
Inset shows a typical $^{13}$C NMR spectrum in field along $c$ direction. 
At high temperature, two peaks labeled as inner and outer sites were observed. 
These peaks were superimposed at low temperature due to line broadening and reduction of Knight shift.  
}
\label{fig1}
\end{figure}

The conventional saturation-recovery method was employed for the $1/T_{1}$ measurement. 
For $\kappa$-(BEDT-TTF)$_{2}X$ salt, when the field is along the $c$ axis, 
two crystallographically independent $^{13}$C sites give rise to two NMR peaks, which are labeled as inner and outer sites \cite{soto-PRB52}. 
At high temperatures, these two peaks can be clearly resolved, as shown in the inset of Fig.~\ref{fig1}, 
and the relaxation time $T_{1}$ can separately be determined. 
As the result, the inner site is found to have three times longer $T_{1}$ than that for outer site. 
This ratio is temperature independent, 
because the difference in $T_{1}$ is determined by the site-dependent, and temperature-independent hyperfine coupling constants. 
At low temperatures, as these peaks are not resolved due to line broadening, and the reduction of Knight shift,  
we integrated the whole spectrum and fitted the obtained recovery curve with two component exponential function,  
\[
\frac{M_0-M(t)}{M_0} = 0.5 \exp \left( \frac{-t}{T_1^{\rm outer}} \right) +0.5 \exp  \left( \frac{-t}{T_1^{\rm inner}} \right).
\]
We fixed the ratio $T_{1}^{\rm inner}/T_{1}^{\rm outer} = 3.$ 
As shown in Fig.~\ref{fig1}, all the recovery curves, including the temperatures and fields above and below the SC transition, 
are sufficiently fitted with this function, which is represented by dashed line. 
The recovery of nuclear magnetization shows a universal behavior, 
when the horizontal axis is scaled by $T_{1}^{\rm inner}$ determined at each temperature. 
This scaling by $T_{1}$ indicates that the obtained $T_{1}$ characterize correctly the time scale for nuclear magnetization to relax, 
regardless of the fitting procedures. 

We also performed NMR experiments on deuterated $\kappa$-$d$[$n,n$]-Br salts ($n=2,4$). 
For the deuterated samples, field orientation was fixed to the $c$ direction to eliminate the angle dependence of $1/T_{1}T$ values 
originating from the angle dependent hyperfine coupling constant and investigated quantitatively the deuteration effect.

\section{resultS and discussions}
\begin{figure}[tbp]
\begin{center}
\includegraphics[width=8cm]{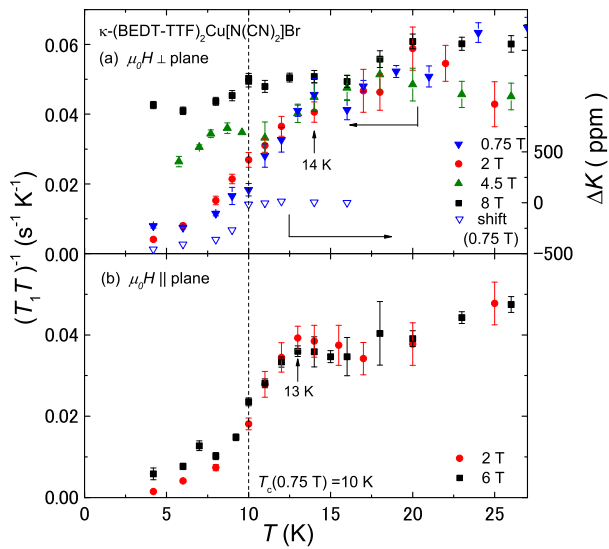}
\end{center}
\caption{ 
Temperature dependence of $1/T_{1}T$ for $\kappa$-Br salt in various magnetic fields perpendicular (a) and parallel (b) to the conduction plane. 
Open symbol shows the temperature dependence of NMR shift in $0.75$ T. 
Dashed line denotes the mean-field $T_{c} = 10$ K, which was determined by the abrupt decrease in NMR shift. 
The reduction of $1/T_{1}T$ observed above $T_{c}$ indicates that the fluctuating SC order parameter affects the quasi-particle DOS. 
}
\label{fig2}
\end{figure}

Figure~\ref{fig2} shows the temperature dependence of $1/T_1T$ for $\kappa$-Br salt in various magnetic fields. 
The temperature-independent Fermi liquid behavior was observed below $20$~K, 
which confirms that the magnetic fluctuations near antiferromagnetic transition 
are sufficiently suppressed, and the Fermi liquid state is established when superconductivity sets in. 
In the Fermi liquid state, $1/T_{1}T$ is proportional to the square of the DOS at the Fermi energy $N(E_F)$, 
\begin{equation}
\frac{1}{T_{1}T} = \frac{\pi k_B}{\hbar}A_{\rm hf}^2(\theta )N^2(E_{F}).
\end{equation}
Here, $A_{\rm hf}(\theta)$ is the angle dependent hyperfine coupling constant. 
This Korringa relation is modified either by enhanced magnetic fluctuations near magnetic transition, or by
the SC fluctuation near $T_{c}$, which corrects the quasi-particle DOS \cite{carretta-PRB61, kuboki-JPSJ58}. 
The vertical dashed line in Fig.~\ref{fig2} denotes the mean-field $T_{c}$ at $0.75$ T ($T_{c}(0.75~{\rm T})=10$~K), 
which was determined by the abrupt decrease in NMR shift at $0.75$ T. 
(Right scale of Fig.~\ref{fig2}(a))
When magnetic field of $8$ T is applied perpendicular to the conduction plane $H_{\perp}$ (Fig.~\ref{fig2}(a)), 
superconductivity is completely destroyed and Fermi liquid behavior persists down to $4.2$~K. 
In smaller fields, $1/T_{1}T$ deviates from the Fermi liquid behavior below $14$~K, which is higher than the mean-field $T_{c}$ of $10$ K. 
We suggest that the decrease in $1/T_{1}T$ above $T_c$ is caused by the fluctuating SC order parameter.

The reduction of $1/T_{1}T$ was also observed in the parallel magnetic field (Fig.~\ref{fig2}(b)). 
As $6$ T is not strong enough to suppress superconductivity and induce Fermi liquid behavior at low temperatures, 
the onset of $1/T_1T$ reduction was determined as the temperature at which $1/T_{1}T$ deviates from the high-temperature Fermi liquid behavior. 
The onset for $H_{\perp}$ ($14$~K) is slightly higher than that for $H_{\parallel}$ ($13$~K). 
This subtle difference is due to the anisotropy of coherence length. 
Since decrease in $1/T_{1}T$ at temperatures higher than $T_{c}$ was observed in both $H_{\perp}$ and $H_{\parallel}$, 
we conclude that $\kappa$-Br salt shows fluctuating SC state regardless of the magnetic field directions.
Theoretical study has revealed that SC fluctuations can correct quasi-particle DOS in three different processes, 
which are called Aslamazov-Larkin (AL), Maki-Tompson (MT), and DOS processes \cite{larkin}. 
The AL process is irrelevant to the dynamical susceptibility, which determines $1/T_{1}T$, 
Present results indicate that the DOS process, which reduces the quasi-particle DOS, is dominant in $\kappa$-Br salt 
compared to the MT process, which increases $1/T_{1}T$.

\begin{figure}[tbp]
\begin{center}
\includegraphics[width=8cm]{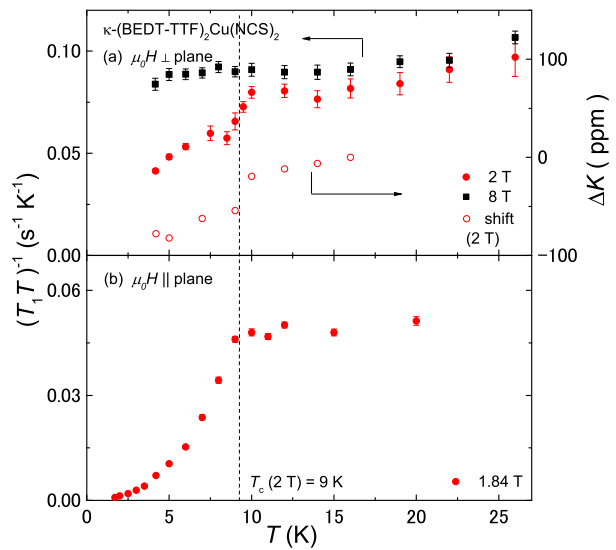}
\end{center}
\caption{
Temperature dependence of $1/T_{1}T$ for $\kappa$-NCS salt in magnetic fields perpendicular (a) and parallel (b) to the conduction plane. 
Open symbol shows the temperature dependence of NMR shift in $2$ T, which determined the mean-field $T_{c} = 9$ K. 
Although $T_{c}$ is comparable to that for $\kappa$-Br salt, 
$\kappa$-NCS salt shows the reduction of $1/T_{1}T$ above $T_{c}$ only in the temperature range very close to $T_{c}$. 
}
\label{fig3}
\end{figure}

The same measurements were performed for $\kappa$-NCS salt, and the results were displayed in Fig.~\ref{fig3}. 
The Fermi liquid behavior was observed at all temperature region in $H_{\perp} = 8$ T. 
At a small field, deviation of $1/T_{1}T$ from the Fermi liquid behavior starts almost at the mean-field $T_{c}$ at $2$ T ($T_{c}(2~{\rm T}) = 9$~K). 
Similarly, in a parallel field, the decrease in $1/T_{1}T$ from the temperature-independent value 
was observed only below $T_{c}$. 
These results indicate that the temperature region, where SC fluctuations become significant, 
is limited very close to $T_{c}$ in $\kappa$-NCS salt. 

Now, we compare the results of $\kappa$-Br and $\kappa$-NCS salts. 
Although SC transition occurs at comparable temperatures in both salts, 
the reduction of $1/T_{1}T$ due to the SC fluctuation was observed only in $\kappa$-Br salt, 
in consistent with Nernst effect and magnetization measurements \cite{nam-nature449, uehara-JPSJ82}. 
The SC parameters for $\kappa$-Br and $\kappa$-NCS salts should be compared to understand the 
universal behavior of fluctuating SC state.

\begin{figure}[tbp]
\begin{center}
\includegraphics[width=7.5cm]{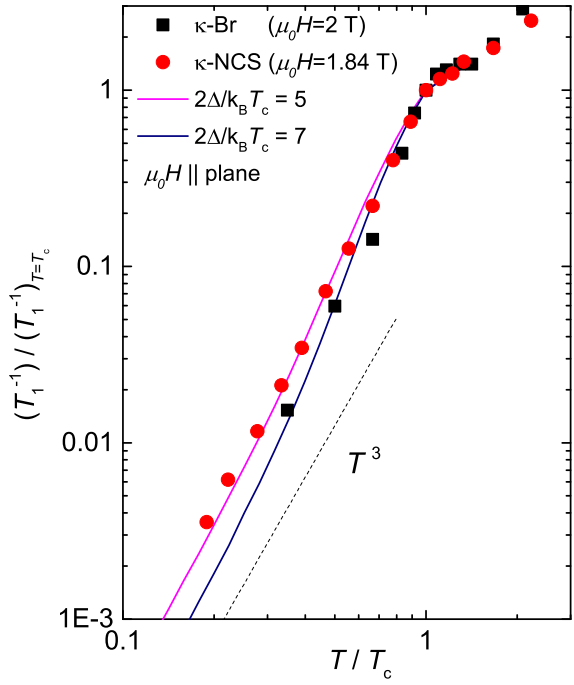}
\end{center}
\caption{
SC parameter $2\Delta/ k_{B} T_{c}$ estimated from the temperature dependence of $1/T_{1}$ in the SC state. 
The vertical axis is $1/T_{1}$ scaled by the value at $T_{c}$, and the horizontal axis is the reduced temperature $T/T_{c}$. 
We estimated $2 \Delta / k_{B} T_{c}$ by fitting the experimental data to the theoretical curve for the $d$-wave superconductor. 
}
\label{fig4}
\end{figure}

In the critical regime above $T_{c}$, the short-range SC coherence can be observed 
when the thermal energy becomes comparable to the energy cost required to induce SC gap within the Pippard length $\xi$. 
Therefore, the critical temperature range is determined by the SC gap size $\Delta$, $T_{c}$, and $\xi$, as 
\begin{equation} \label{eq1}
\frac{|T-T_{c}|}{T_{c}} \equiv G \propto \left( \frac{k_BT_c}{\xi_{\parallel}^2 \xi_{\perp} \Delta} \right)^2, 
\end{equation}
where $\xi_{\parallel}$ and $\xi_{\perp}$ are the in-plane and inter-plane Pippard length. 
The experimentally determined Pippard lengths are $\xi_{\parallel}=3.7$ nm and $\xi_{\perp}=0.6$ nm for $\kappa$-Br, and 
$\xi_{\parallel}=7.0$ nm and $\xi_{\perp}=0.5$ nm for $\kappa$-NCS \cite{dressel-PRB50}.
The typical temperature range for a three-dimensional BCS superconductor is as narrow as $G \sim 10^{-6}$. 
Wider critical regime is expected for low dimensional superconductors with short $\xi_{\perp}$. 
Our experimental results suggest that $G$ of $\kappa$-Br salt reaches to the order of $10^{-1}$, 
which is comparable to the results obtained for high-$T_{c}$ cuprate.

We estimated the ratio $2\Delta/k_BT_{c}$ from the temperature dependence of $1/T_{1}$ in the SC state. 
Figure~\ref{fig4} shows the temperature dependence of $1/T_1$ normalized at $T_{c}$ for $\kappa$-Br and $\kappa$-NCS salts. 
The power-law temperature dependence suggests that 
$d$-wave superconductivity with nodes on SC gap is realized in both $\kappa$-Br and $\kappa$-NCS salts. 
The fit to the temperature dependence with the theoretical curve for $d$-wave superconductivity gives 
the estimation of $2\Delta/k_{B}T_{c}$, which are $7$ for $\kappa$-Br salt and $5$ for $\kappa$-NCS salt. 
The ratio between $\kappa$-Br and $\kappa$-NCS salts is in perfect agreement with the values reported in the literature \cite{dressel-PRB50}. 
We note that the large $\Delta$ for $\kappa$-Br salt seems to reduce critical temperature regime defined in eq.~(\ref{eq1}). 
However, because the Pippard length is also related to $\Delta$ as 
$\xi = \hbar v_{F}/\pi \Delta$, the larger $\Delta$ increases the critical regime by reducing the coherence length. 
Here we assumed that only the coherence length in the conduction plane $\xi_{\parallel}$ is determined by $\Delta$ 
because of the two-dimensional nature of superconductivity in $\kappa$ salts. 
In fact, $\xi_{\parallel}$ for $\kappa$-Br salt is twice shorter than that for $\kappa$-NCS salt, 
whereas $\xi_{\perp}$ is almost comparable between $\kappa$-Br and $\kappa$-NCS salts \cite{dressel-PRB50}.

\begin{figure}[tbp]
\begin{center}
\includegraphics[width=8cm]{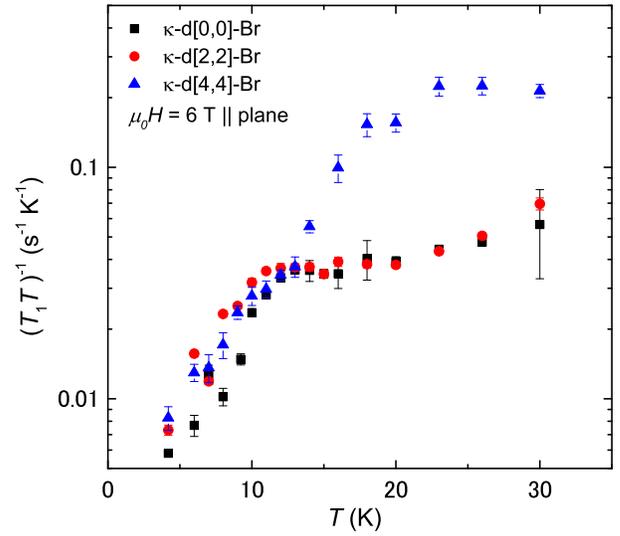}
\end{center}
\caption{
Deuteration effect on the FL state near $T_{c}$. 
Strong temperature dependence due to the reduction of spin term was observed in $\kappa$-$d[4,4]$-Br salt, 
while $\kappa$-$d[0,0]$-Br and $\kappa$-$d[2,2]$-Br salts show the FL liquid behavior below $20$ K. 
In $\kappa$-$d[4,4]$-Br, the effects of the SC fluctuation was contaminated by the 
strong magnetic fluctuations near the antiferromagnetic phase transition. 
}
\label{fig5}
\end{figure}

The larger $\Delta$ is expected for the deuterated $\kappa$-$d[4,4]$-Br salt, 
which locates very close to the antiferromagnetic phase. 
Previous NMR study for $\kappa$-$d[4,4]$-Br salt suggested a possibility 
that the critical SC regime starts from rather high temperatures \cite{miyagawa-PRL89}. 
We also measured $1/T_{1}T$ near $T_{c}$ for $\kappa$-$d[n,n]$-Br salts with $n=0,2,4$ applying the magnetic field of $6$ T along the $c$ direction. 
In Fig.~\ref{fig5}, a strong temperature dependence of $1/T_{1}T$ was observed around $20$~K in $\kappa$-$d[4,4]$-Br salt 
in consistent with previous report \cite{miyagawa-PRL89}. 
This behavior cannot be directly interpreted as the effect of fluctuating SC order parameter, 
because enhanced magnetic fluctuations contaminate the pure effect of SC fluctuations. 
In the vicinity of antiferromagnetic phase transition, 
the observed $1/T_{1}T$ is the sum of the contributions from weakly correlated Fermi surface (FL term) and 
strongly nested Fermi surface at the ordering vector $\bm{Q}$ (spin term), 
\begin{equation}
\frac{1}{T_{1}T} = \left( \frac{1}{T_{1}T} \right)_{\rm FL} + \left( \frac{1}{T_{1}T} \right)_{\rm spin}. 
\end{equation}
The strong temperature dependence observed in $\kappa$-$d[4,4]$-Br salt 
indicates that the spin term remains dominant until low temperatures. 
The increase in $1/T_{1}T$ by enhanced magnetic fluctuations was observed only above $30$ K 
in $\kappa$-$d[0,0]$-Br and $\kappa$-$d[2,2]$-Br salts.
This spin term is suppressed below $20$ K, and the Fermi liquid state is stabilized at $T_{c}$ in these salts. 
The comparable constant value of $1/T_{1}T = 0.035$~s$^{-1}$K$^{-1}$ indicates that 
the deuteration does not significantly modify the FL term. 
We suggest, from the deuteration independent behavior of $1/T_{1}T$ below $14$ K, that 
$\kappa$-$d[4,4]$-Br salt also shows the reduction of FL term due to SC fluctuation near $T_{c}$, 
although details were hindered by the spin term. 

\section{Conclusion}
In conclusion, 
we observed the reduction of $1/T_{1}T$ starting above $T_c$ in $\kappa$-Br salt in fields both parallel and perpendicular to the conduction plane, 
and addressed the fluctuating SC order parameter as the origin of this reduction. 
The microscopic NMR experiment enabled the direct observation of quasi-particle DOS in the fluctuating SC state. 
We revealed that, in the fluctuating SC state in $\kappa$-Br salt, 
the quasi-particle DOS is reduced from the normal-state value because of the dominant DOS process. 
In $\kappa$-NCS salt, the reduction of quasi-particle DOS due to SC fluctuation was observed only in the temperature range very close to $T_{c}$. 
Our systematic study clarified the field-orientation dependence of the SC fluctuation 
and convinced that the SC fluctuation has larger effect on $1/T_{1}T$ in $\kappa$-Br salt than that in $\kappa$-NCS salt. 
From the estimation of SC gap size for $\kappa$-Br and $\kappa$-NCS salts, 
we suggest that the small critical regime in $\kappa$-NCS salt is due to the long coherence length. 
We also measured $1/T_{1}T$ in the deuterated $\kappa$-$d[n,n]$-Br salts, 
and found that the deuteration induces the large spin term even at $T_{c}$, which contaminates the pure effect of SC fluctuation. 
These results indicate that $\kappa$-$d[0,0]$-Br salt is an ideal system to study pure effect of SC fluctuation, and 
provide microscopic information to understand the electronic properties in the fluctuating SC state. 

We acknowledge Ryusuke Ikeda for fruitful discussion. 
This work is partially supported by Grant-in-Aid for Challenging Exploratory Research (No. 25610082).


\begin{thebibliography}{99}

\bibitem{millis-PRB48}
A.~J.~Millis, 
Phys. Rev. B {\bf 48,} 7183 (1993).

\bibitem{xu-nature}
Z.~A.~Xu, and N.~P.~Ong, and Y.~Wang, and T.~Takeshita, and S.~Uchida, 
Nature {\bf 406,} 486 (2000).

\bibitem{larkin}
A.~Larkin, and A.~Varlamov, in {\it Superconductivity}, edited by K. H. Bennemann, and J. B. Ketterson (Springer-Verlag, Berlin, 2008), Vol. 1.

\bibitem{carretta-PRB61}
P.~Carretta, A.~Lascialfari, A.~Rigamonti, A.~Rosso, and A.~Varlamov,
Phys. Rev. B {\bf 61,} 12420 (2000).

\bibitem{mitrovic-PRL82}
V.~F.~Mitrovi\'{c}, H.~N.~Bachman, W.~P.~Halperin, M.~Eschrig, J.~A.~Sauls, A.~P.~Reyes, P.~Kuhns, and W.~G.~Moulton
Phys. Rev. Lett. {\bf 82,} 2784 (1999).

\bibitem{ri-PRB50}
H.-C.~Ri, and R.~Gross, and F.~Gollnik, and A.~Beck, and R.~P.~Huebener, P.~Wagner, and H.~Adrian 
Phys. Rev. B {\bf 50,} 3312 (1994). 

\bibitem{nam-nature449}
M.-S.~Nam, A.~Ardavan, S.~J.~Blundell, and J.~A.~Schlueter,
Nature {\bf 449,} 584 (2007).

\bibitem{mayaffre-PRL75}
H.~Mayaffre, P.~Wzietek, D.~Jer\'{o}me, C.~Lenoir, and P.~Batail,
Phys. Rev. Lett. {\bf 75,} 4122 (1995).

\bibitem{kwok-PRB42}
W.~K.~Kwok, U.~Welp, K.~D.~Carlson, G.~W.~Crabtree, K.~G.~Vandervoort, H.~H.~Wang, A.~M.~Kini, J.~M.~Williams, D.~L.~Stupka, L.~K.~Montgomery, and J.~E.~Thompson,
Phys. Rev. B {\bf 42,} 8686 (1990).

\bibitem{lang-PRB49}
M.~Lang, F.~Steglich, N.~Toyota, and T.~Sasaki,
Phys. Rev. B {\bf 49,} 15227 (1994).

\bibitem{uehara-JPSJ82}
T.~Uehara, M. ~Ito, H.~Taniguchi, and K.~Satoh,
J. Phys. Soc. Jpn. {\bf 82,} 073706 (2013).

\bibitem{tsuchiya-PRB85}
S.~Tsuchiya, J.~I.~Yamada, S.~Tanda, K.~Ichimura, T.~Terashima, N.~Kurita, K.~Kodama, and S.~Uji,
Phys. Rev. B {\bf 85,} 220506 (2012).

\bibitem{randeria-PRB50}
M.~Randeria, and A.~A.~Varlamov,
Phys.~Rev.~B {\bf 50,} 10401 (1994).

\bibitem{urayama-chemlett17}
H.~Urayama, H.~Yamochi, G.~Saito, K.~Nozawa, T.~Sugano, M.~Kinoshita, S.~Sato, K.~Oshima, A.~Kawamoto, J.~Tanaka, and T.~Mori,
Chem. Lett. {\bf 17,} 55 (1988).

\bibitem{welp-physicaB186-188}
U.~Welp, S.~Fleshler, W.~K.~Kwok, G.~W.~Crabtree, K.~D.~Carlson, H.~H.~Wang, U.~Geiser, J.~M.~Williams, and V.~M.~Hitsman,
Physica B {\bf 186-188,} 1065 (1993).

\bibitem{kawamoto-PRL74}
A.~Kawamoto, K.~Miyagawa, Y.~Nakazawa, and K.~Kanoda, 
Phys.~Rev.~Lett. {\bf 74,} 3455 (1995).

\bibitem{strack-PRB72}
Ch.~Strack, C.~Akinci, V.~Paschenko, B.~Wolf, E.~Uhrig, W.~Assmus, M.~Lang, J.~Schreuer, L.~Wiehl, J.~A.~Schlueter, J.~Wosnitza, D.~Schweitzer, J.~M\"{u}ller, and J.~Wykhoff, Phys.~Rev.~B {\bf 72,} 054511 (2005).

\bibitem{kawamoto-JACS120}
A.~Kawamoto, H.~Taniguchi, and K.~Kanoda,
J. Am. Chem. Soc. {\bf 120,} 10984 (1998).

\bibitem{miyagawa-PRL89}
K.~Miyagawa, A.~Kawamoto, and K.~Kanoda,
Phys. Rev. Lett. {\bf 89,} 017003 (2002).

\bibitem{tsuchiya-JPSJ82}
S.~Tsuchiya, J.~Yamada, T.~Terashima, N.~Kurita, K.~Kodama, K.~Sugii, and S.~Uji,
J. Phys. Soc. Jpn. {\bf 82,} 064711 (2013).

\bibitem{yamashita-synthmet133-134}
M.~Yamashita, A.~Kawamoto, and K.~Kumagai,
Synth. Met. {\bf 133-134,} 125 (2003).

\bibitem{pake-JCP16}
G.~E.~Pake,
J. Chem. Phys. {\bf 16,} 327 (1948).

\bibitem{soto-PRB52}
S.~M.~De~Soto, C.~P.~Slichter, A.~M.~Kini, H.~H.~Wang, U.~Geiser, J.~M.~Williams, 
Phys.~Rev.~B, {\bf 52,} 10364 (1995). 

\bibitem{kuboki-JPSJ58}
K.~Kuboki, and H.~Hukuyama,
J. Phys. Soc. Jpn. {\bf 58,} 376 (1989).



\bibitem{dressel-PRB50}
M.~Dressel, O.~Klein, and G.~Gr\"{u}ner, K.~D.~Carlson, H.~H.~Wang, and J.~M.~Williams, 
Phys. Rev. B {\bf 50,} 13603 (1994).



\end{thebibliography}

\end{document}